# Prospective Pathways of Green Graphene-based Lab-on-Chip Devices: The Pursuit towards Sustainability

*Joydip Sengupta*[1] *and Chaudhery Mustansar Hussain*[2*]

[1]Department of Electronic Science, Jogesh Chandra Chaudhuri College, Kolkata - 700033, India

[2]Department of Chemistry and Environmental Science, New Jersey Institute of Technology, USA

*Corresponding author Email: chaudhery.m.hussain@njit.edu

At present, analytical lab-on-chip devices find their usage in different facets of chemical analysis, biological analysis, point of care analysis, biosensors, etc. In addition, graphene has already established itself as an essential component of advanced lab-on-chip devices. Graphene-based lab-on-chip devices have achieved appreciable admiration because of their peerless performance in comparison to others. However, to accomplish a sustainable future a device must undergo "Green-Screening" to check its environmental compatibility. Thus, extensive research is carried out globally to make the graphene-based lab-on-chip green, though it is yet to be achieved. Nevertheless, as a ray of hope, there are few existing strategies that can be stitched together for feasible fabrication of environment-friendly green graphene-based analytical lab-on-chip, and those prospective pathways are reviewed in this paper.

## 1. Introduction

The term lab-on-chip (LOC), coined in 1995 for the first time[1], represents a scaled-down microfluidic device that can assimilate and automate the typical laboratory procedure into a single system with improved speed. Moreover, LOC with an integrated sample preparation facility[2,3] is an ideal one for rapid detection of the sample with low cost and a small quantity of sample consumption. As time progresses, it was found that the inclusion of nanomaterials can appreciably enhance the overall performance of the LOC[4]. Subsequently, graphene was included in the LOC device for the first time in 2011[5]. Since then, the graphene-based lab-on-chip (GLOC) devices have become a matter of keen interest for researchers around the globe[6] and the fabricated devices showed extreme potential in different applications even in the rapid detection of SARS-CoV-2[7,8]. However, in today's world, e-waste has become a matter of major concern as it poses tremendous health and environmental hazard[9]. Thus, to achieve a sustainable future all the LOC devices also must undergo "Green-Screening" to ensure their eco-friendliness. In this regard, various prospective approaches can be adopted by the researchers to fabricate environment-friendly green graphene-based analytical lab-on-chip (GGLOC). The promising pathways that may be coalesced to accomplish the mammoth task of fabricating GGLOC are reviewed here.

## 2. Green pathways for Graphene-based LOC

To achieve GGLOC, each part of the GLOC i.e., the basic structure and the active component (graphene) must be fabricated in an environment-friendly manner. Moreover, the entire microfluidic operation must be carried out also using the green method, i.e., the use of green solvents, green waste management, green power management, etc. (Scheme 1). The green approaches that can be adopted by the researchers for the fabrication of different constituents of the LOC are discussed below.

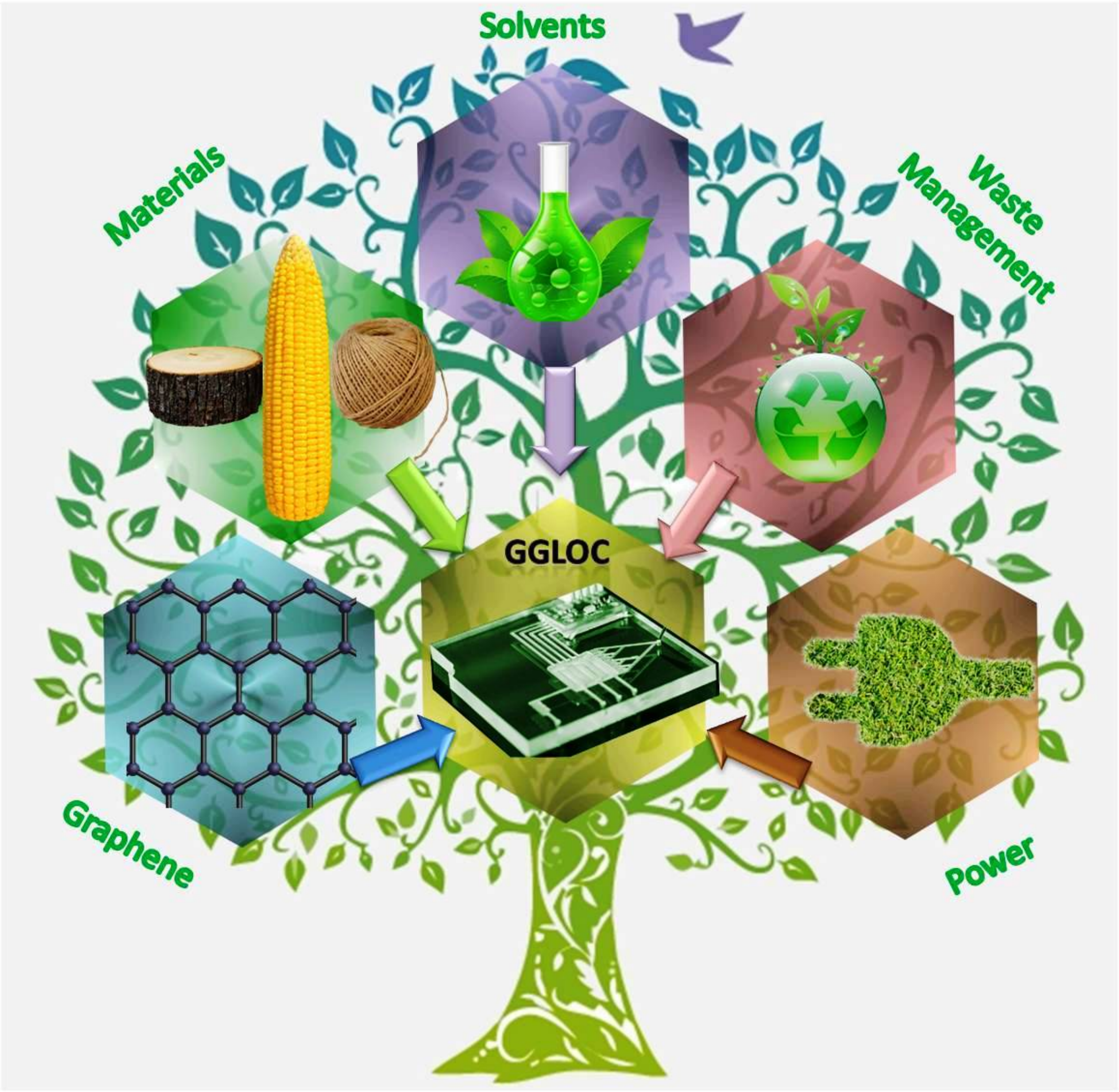

*Scheme 1. The recipe for fabricating GGLOC.*

### 2.1 Green approach for the fabrication of the basic structure of LOC

For the fabrication of the basic structure of LOC, many organic and inorganic materials were used to date[10]. However, most of the materials are not suitable for the fabrication of GGLOC as the material is not environment-friendly and in addition, the material must possess biocompatibility, biodegradability, and renewability. There are some invaluable efforts across the globe to fabricate the basic structure of LOC using eco-friendly green synthesis. The pathway towards green LOC was started probably in the year 2003 through the initiative of Grayson et al.[11]. They fabricated biodegradable[12] polymeric microchips employing poly (L-lactic acid) (PLLA). 36 reservoirs were housed in the 480-560 µm thick microchips having a diameter of 1.2 cm and each of the reservoirs can potentially be filled with a different chemical (Fig 1).

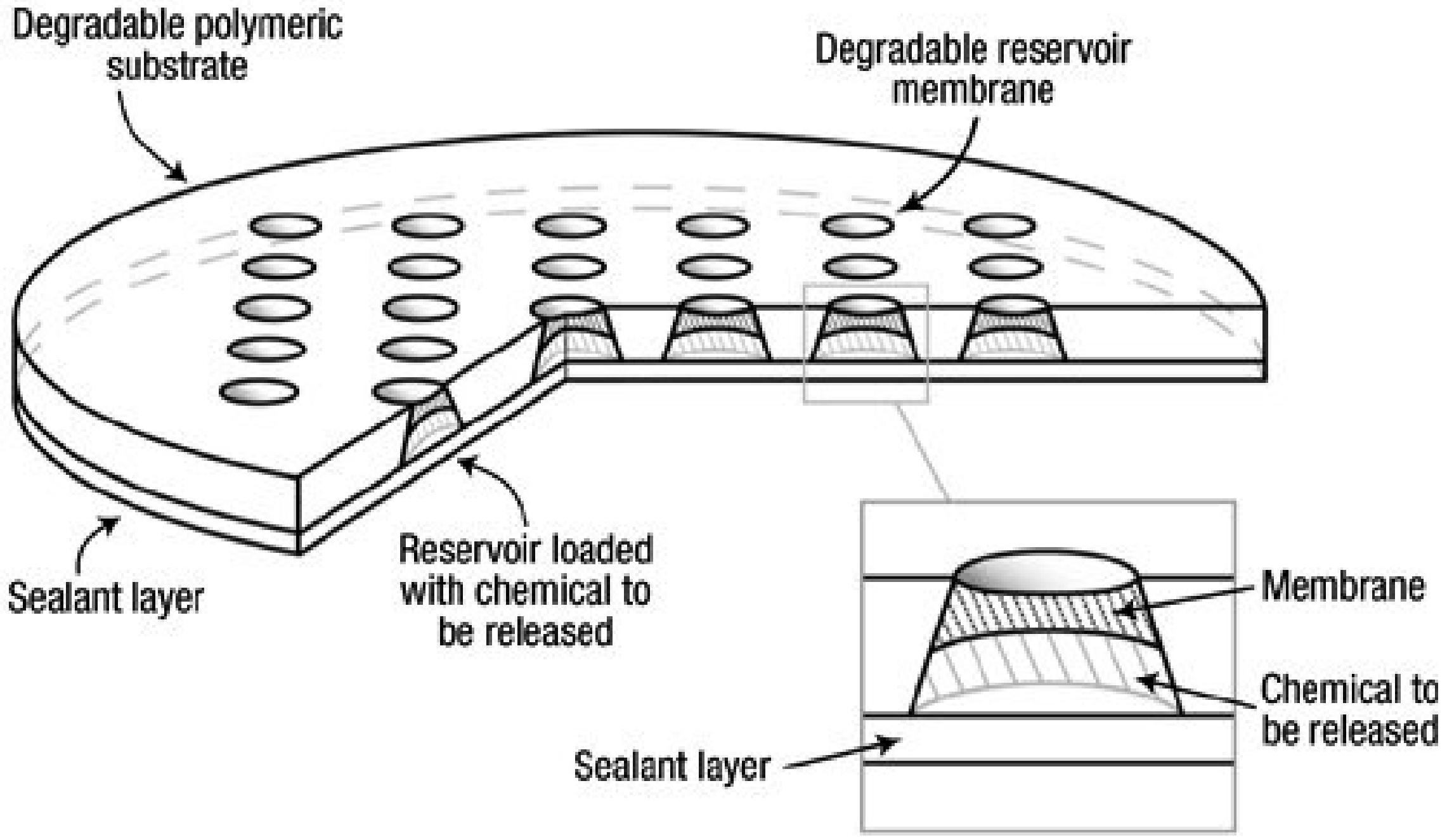

*Fig 1. Diagram of polymeric microchip device. The main body of the device is composed of a reservoir-containing substrate that is fabricated from a degradable polymer. Truncated conical reservoirs in the substrate are loaded with the chemical to be released and sealed with polymeric degradable reservoir membranes on one end and a sealant layer (polyester tape) on the opposite end. Inset, close-up of a reservoir, reservoir membrane, sealant layer, and chemical to be released.* (Reproduced with permission from Nature Materials 2, no. 11 (November 2003): 767–72, https://doi.org/10.1038/nmat998)

In 2004 King et al.[13] introduced biodegradable thermoplastic poly (DL-lactic-co-glycolide) (PLGA) for the fabrication of microfluidic devices with high resolution, having µm size features. Melt processing and thermal fusion bonding processes were adopted to overcome the problem of bubble creation, film shrinkage, and finally to create a 3D structure. They finally deployed it for successful therapeutic applications. In the very next year, Cabodi et al.[14] fabricated a microfluidic system using soft lithography and housed it within calcium alginate which is known for its low cytotoxicity, biodegradability, and environmental friendliness. They demonstrated that the fabricated LOC system can be used to develop microfluidic scaffolds for tissue engineering. Poly (glycerol sebacate) (PGS), a biocompatible and eco-friendly elastomer having excellent mechanical property, was chosen by Bettinger et al.[15] for the fabrication of a three-dimensional microfluidic tissue-engineering scaffold. They demonstrate cell viability and function using a model hepatocyte cell line and proposed that the LOC can potentially be integrated into the vasculature of the patient to restore organ function. The biological enzymatic crosslinking process was adopted by Paguirigan et al.[16] to fabricate cytocompatible microfluidic devices from gelatine which can be used for adherent cell culture and analysis. Moreover, low autofluorescence of gelatine provides the opportunity to use fluorescence to visualize the 3-D structures which in turn will facilitate additional analysis of structures and allocations of different cell types. Biodegradable silk fibroin microtubes were prepared by Kaplan et al.[17] by dipping straight stainless-steel wire into aqueous silk fibroin (derived from Bombyx mori silkworm cocoons) and the porosity was controlled by adding poly (ethylene oxide) (PEO). The synthesized microtubes can be used to repair blood vessels at the microvascular scale. Authors also varied the porosity of the microtubes and studied the pros and cons of the same based on several parameters such as mechanical strength, protein permeability, enzymatic degradation, cellular permeability, etc. Agarose is another environment-friendly polymer with sufficient mechanical strength in comparison to commonly used natural polymers. Agarose was used by Ling et al.[18] to fabricate cell-laden agarose microfluidic devices by standard soft lithographic techniques. Completely sealed channels with dissimilar aspect ratios were reproduced with high reliability, which demonstrated the suitability of agarose as a potential green material for microfluidic applications. Martinez et al.[19] exhibited a simple, economical yet exceptional technique by using paper as a platform for running multiplexed bioassays. They used chromatography paper to create millimetre-sized channels using standard lithography techniques for the simultaneous detection of glucose and protein in 5 ml of urine. Later on, this group developed paper-based 3D microfluidic devices[20] (Fig 2) for electrochemical sensing[21].

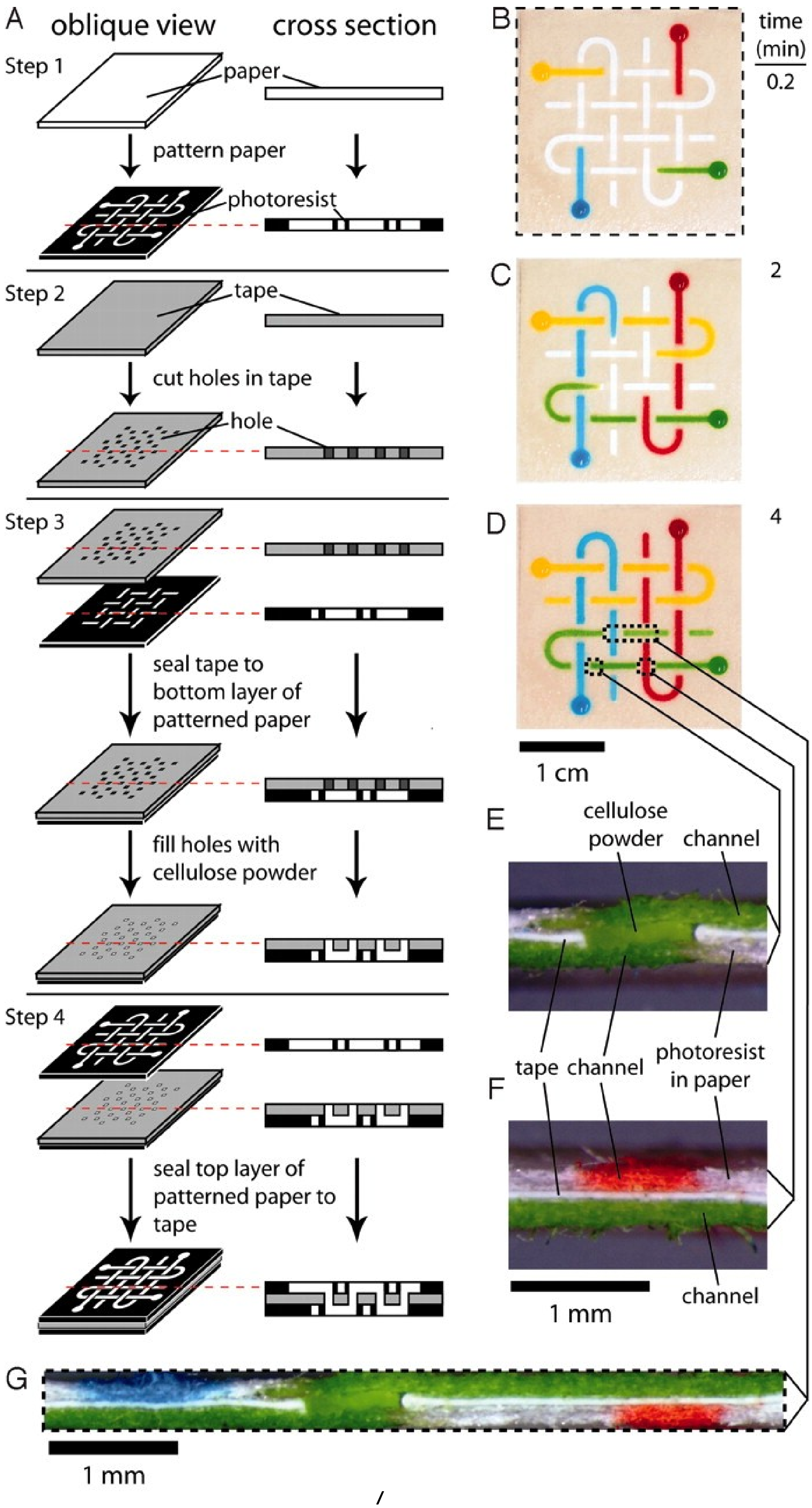
A
oblique view
cross section
Step 1
paper
pattern paper
photoresist
Step 2
tape
cut holes in tape
hole
Step 3
seal tape to bottom layer of patterned paper
fill holes with cellulose powder
Step 4
seal top layer of patterned paper to tape
B
time (min)
0.2
C
2
D
4
1 cm
E
cellulose powder
channel
tape
channel
photoresist in paper
F
channel
1 mm
G
1 mm

/

*Fig 2. Preparation and demonstration of a 3D μPAD. (A) Fabrication. (B) Photograph of a basket-weave system 10 s after adding red, yellow, green, and blue aqueous solutions of dyes to the sample reservoirs. The dotted lines indicate the edge of the device. (C and D) Photographs were taken 2 (C) and 4 (D) min after adding the dyes. The streams of fluids crossed each other multiple times in different planes without mixing. The dotted lines in D show the positions of the cross-sections shown in E, F, and G. (E) Cross-section of the device showing a channel connecting the top and bottom layers of paper. (F) Cross-section of the device showing the three layers of the device with orthogonal channels in the top and bottom layers of paper. (G) Cross-section of the device showing the layers and the distribution of fluid (and colours) in each layer of the device shown in D. The dotted lines indicate the edges of the cross-section.* (Reproduced with permission from Proceedings of the National Academy of Sciences 105, no. 50 (16 December 2008): 19606–11, https://doi.org/10.1073/pnas.0810903105)

Another breakthrough was registered in 2011 when fabric was introduced as a platform for microfluidic device Bhandari et al.[22] by employing silk yarns and simple weaving. They demonstrated that fabrication of complex patterns having reagents can be performed with site-selectivity through employing handlooms. They further developed woven electrochemical fabric-based test sensors[23] and exhibited the potential of their device as sensitive surface-enhanced Raman spectroscopy (SERS) substrates for analytical applications[24]. Vegetable such as corn was used by Luecha et al.[25] for the fabrication of green microfluidic devices. The corn protein (zein) was extracted from protein-rich "distillers' dried grains with soluble" (DDGS) and good quality film can be formed using zein owing to its unique thermoplastic property. Thin zein films containing microfluidic chambers and channels were fabricated by standard soft lithography and stereolithography techniques. The fabricated LOC device displayed adequate strength to aid fluid flow in a complex microfluidic design without any leakage even under high hydraulic pressure. Wei et al.[26] introduced a new idea by exhibiting that, for sample separation on a LOC device, thin polyester threads of various diameters can be used instead of liquid channels. They employed plasma treatment to augment the wettability and surface quality which resulted in a ten-time enhancement in electrochemical detection. Flexibility is an important parameter for any device structure and in the case of LOC, it's more important as LOCs are also used as point of care (POC) devices. Researching towards this direction, Cabrera et al.[27] developed flexible microfluidic devices using natural rubber for optical and electrochemical applications. Specifically, they used latex collected from trees of Hevea Brasiliensis and the fabricated LOC is not only flexible, but transparent also, which makes it appropriate for spectroscopic measurements in the visible range. They used the normal casting method using an indigenous fabrication mould and followed the thermal treatment and demoulding to give the final shape to the LOC device. A natural resinous product named Shellac was used by Lausecker et al.[28] to fabricate a green LOC device. Shellac is a natural thermoplastic with high chemical stability and can form smooth surfaces at low temperatures. The researchers used hot embossing of natural shellac for green fabrication of inexpensive POC device with high replication accuracy, incorporating an energy-efficient procedure where all the consumables are renewable materials. Andar et al.[29] introduced another facile technique for the fabrication of green environment-friendly LOC for POC application using wood as a fabrication material. They used laser engraving and traditional mechanical methods for the development of LOC structures and employed precise surface coatings to overcome the wicking effect of wood. They did several experiments successfully as proof of concept such as rapid protein detection etc. Speller et al.[30] reported a green technique for rapid prototyping of a 3D polydimethylsiloxane (PDMS)-based microfluidic structure using white glue with minimum widths of 200 μm and adjustable heights. A tabular representation depicting the historical evolvements of green materials for fabricating the basic structure of GGLOC is shown in Table 1. In summary, since the last two decades, many green materials were introduced in chip devices; however, the ready availability of basic manufacturing material at an economical price remains a principal factor in determining the most suitable one. In that respect, paper, fabric, corn, rubber, and wood are available in plenty. Moreover, most of the POC devices are now of wearable class. From this perspective, the fabric may be the best choice as a green material for the fabrication of graphene-based POC devices in the near future.

Table 1. Historical evolvements of green materials for fabricating the basic structure of GGLOC

| **Green Material** | **Year of Introduction in Chip Devices** | **Reference** |
|---|---|---|
| Poly (L-lactic acid) (PLLA) | 2003 | [12] |
| Poly (DL-lactic-co-glycolide) (PLGA) | 2004 | [13] |
| Calcium alginate | 2005 | [14] |
| Poly (glycerol sebacate) (PGS) | 2006 | [15] |
| Gelatine | 2006 | [16] |
| Silk fibroin | 2007 | [17] |
| Agarose | 2007 | [18] |
| Paper | 2007 | [19] |
| Fabric | 2011 | [22] |
| Corn | 2011 | [25] |
| Polyester threads | 2013 | [26] |
| Rubber | 2014 | [27] |
| Shellac | 2016 | [28] |
| Wood | 2019 | [29] |

## 2.2 Green Synthesis of Graphene

Carbon is the magical material wherefrom a new branch of nanoscience called "carbon nanoscience" has evolved, stepping on fullerene, carbon nanotube, and lastly graphene. In 2004, the facile technique of making graphene was reported by Novoselov et al.[31] and since then graphene has been the key ingredient in most of the advanced recipes of making nanodevices. The peerless interest in graphene originated from its novel structural, physical, and chemical properties. Structurally graphene can be considered as two dimensional, atomically thin, single layer of graphite with $sp^2$ hybridization and high transparency. It possesses exceptional electrical conductivity owing to its $\pi$ bond located vertically in the lattice plane. Bandgap engineering is also feasible with graphene by modifying the width of graphene nanoribbon. The super-light graphene with a high specific surface area, also exhibits high thermal conductivity and mammoth mechanical strength. Thus, graphene, with exotic electrical, mechanical, and thermal properties, is the obvious choice for the material scientist for exploitation in LOC devices.

Graphene being the main component of GLOC, green synthesis of graphene is of utmost importance. To date, numerous pathways were chosen for the synthesis of graphene using the green procedure. The journey started in 2008 when Fan et al.[32] used deoxygenation of exfoliated graphite oxide (GO) under alkaline conditions for the first green synthesis of graphene at low temperatures without requiring reducing agents (Fig 3).

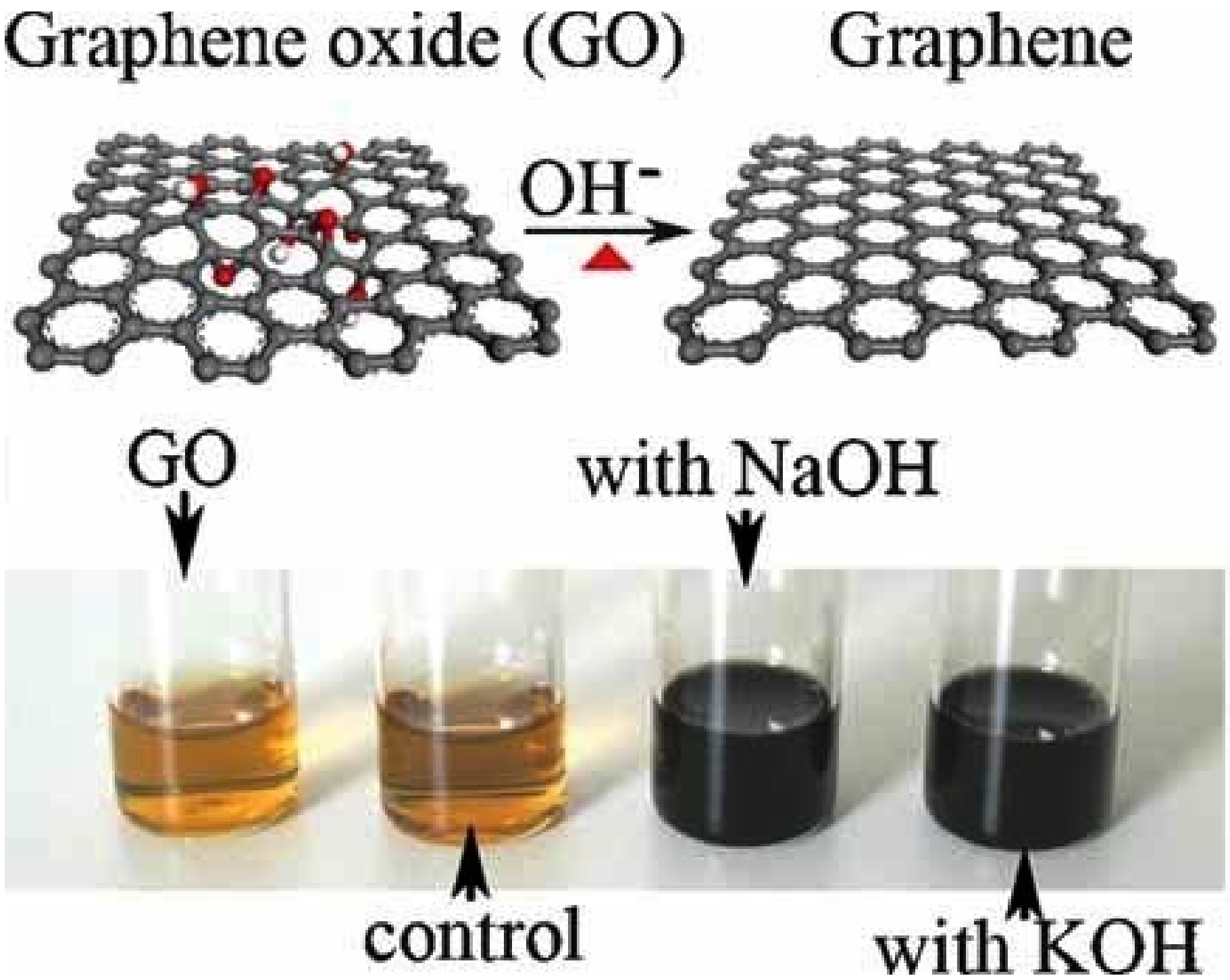

*Fig 3. (above) Illustration for the deoxygenation of exfoliated GO under alkaline conditions and (below) images of the exfoliated GO suspension (0.5 mg m/L) before and after the reaction. The control experiment (below) was carried out by heating the pristine exfoliated-GO suspension without NaOH and KOH at 90°C for 5 h with the aid of sonication. Note that no obvious colour change is observed in the control, even after prolonged heating at relatively high temperatures.* (Reproduced with permission from Advanced Materials 20, no. 23 (2008): 4490–93, https://doi.org/10.1002/adma.200801306)

Guo et al.[33] devised a method to prepare a high-quality environment-friendly green graphene having good electrical conductivity by electrochemical reduction of exfoliated GO precursor. They found some defects in the graphene sheet originated from the rapid reduction; however, annealing can be used to heal the defects. Inexpensive source material with high availability always facilitates the synthesis of nanomaterials at the industrial scale. In this regard, Zhu et al.[34] devised a green method to prepare graphene by reducing the sugar by employing exfoliated GO precursor. The synthesized graphene showed good stability in water for more than one month, which indicates that this facile approach can be used for the large-scale production of water-soluble graphene. Reduction of GO to prepare graphene often employs harmful chemicals. Fan et al.[35] developed a green and inexpensive method to produce graphene by Fe reduction of exfoliated GO. The Fe can be easily removed from obtained graphene by magnetic separation. Yi et al.[36] presented a facile and green synthesis of graphene by liquid-phase exfoliation of graphite. They exhibited that simple sonication of a mixture containing graphite powder, water and alcohol can result in high-quality graphene nanosheets with a yield of almost10 wt%. Furthermore, nearly 86% of the obtained graphene was less than 10-layer thick with a monolayer fraction of approximately 8%. An environment-friendly, electrochemical approach was adopted by Liu et al.[37] to prepare graphene flakes employing graphite rods with a yield of nearly 50%. They reported that the synthesized green graphene has excellent electrochemical properties, thus can be used as anode material for lithium-ion batteries. Carrot extracted anti-oxidants (carotenoids) were used by Vusa et al.[38] to prepare green graphene. They employed carotenoids namely β-carotene, lutein, and lycopene, for the reduction of GO to obtain graphene (Fig 4).

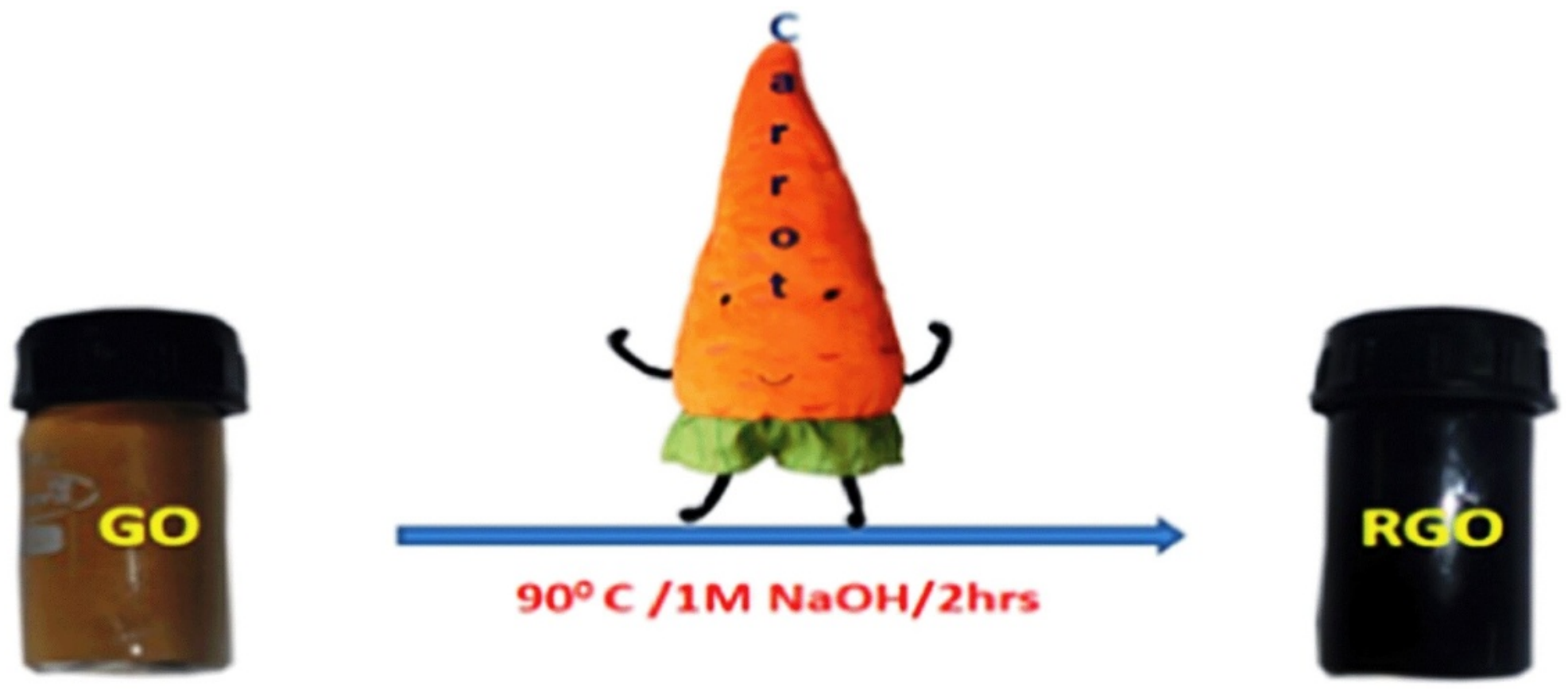

*Fig 4. Schematic representing the reduction of GO by carrot.* (Reproduced with permission from RSC Advances 4, no. 43 (23 May 2014): 22470–75, https://doi.org/10.1039/C4RA01718H)

Xu et al.[39] introduced a few more green reductants for the synthesis of graphene. Three different environment-friendly green reductants namely L-ascorbic acid (L-AA), D-glucose (D-GLC), and tea polyphenol (TP) were used for the chemical reduction of GO. They reported that L-ascorbic acid can produce the best quality graphene with electrical conductivity (9.8 $S.cm^{-1}$) among all the obtained graphene samples. The grape seed extract was used for the synthesis of graphene by Yaragalla et al.[40] employing a green approach. GO powder was mixed with grape seed extract in a 1:2 ratio for the reduction of GO to form graphene with an average thickness of 4.2 nm and bandgap 3.84 eV. The obtained graphene also exhibited efficient anti-proliferative and antimicrobial activity. A biomolecule named alanine was used by Wang et al.[41] for the one-step environment-friendly green reduction of GO to produce graphene on a large scale without using any stabilizer or alkaline medium. They varied treatment time, temperature, and concentration of the reagent to optimize the graphene synthesis procedure. A facile water-based approach was used by Ding et al.[42] to prepare the aqueous compatible environment-friendly graphene on a large scale. They adopted a liquid exfoliation route to produce graphene from natural graphite in pure water with the assistance of vapour pre-treatment, not including any chemicals or surfactants. The obtained graphene has fewer than ten atomic layers and can be stored in dispersed form. Abdullah et al.[43] prepared few-layered graphene using bath sonication of graphite. They used coffee as a liquid phase exfoliating medium for direct synthesis of environment-friendly graphene by graphite exfoliation. The bandgap of the obtained graphene is 3.1 eV with a thickness of 2 to 4 layers. Green synthesis of few-layer graphene from graphite was carried out by Gürünlü et al.[44] in a molten salt mixture of LiCl-KCl. The carbonization temperature varied from 500 to 800°C to study its effect on the obtained graphene. The environment-friendly graphene grown at 600°C showed the highest conductivity value of 1219 S/m. A one-pot preparation method was reported by Hwang et al.[45]for the synthesis of eco-friendly reduced graphene oxide quantum dots. They used perfluorotributylamine solution consisting of three butyl fluoride groups connected to an amine centre able to produce reduced-graphene oxide quantum dots at 160 °C from reduced-graphene oxide precursor. The bandgap of the quantum dots can be varied by changing the temperature of the solution. Bacillus sphaericus is a Gram-positive aerobic bacterium that is commonly found in soil. Bacillus sphaericus was used by Xu et al.[46] for the reduction of GO to obtain environment-friendly graphene with good conductivity (0.8 to 1.1 mW·cm$^{-2}$). Along with some less recognized chemicals and bacteria, a few well-known eatables like sugar, carrot and grape were used for the green synthesis of graphene. In general, the cost of the product is proportionate with the cost of its raw material. Thus, it can be opined that for the green synthesis of graphene researchers must strive to employ common, inexpensive raw materials to make green graphene cheaper.

### 2.3 Green Solvents

There are two routes to ensure the efficient use of solvents in order to save the environment, either by employing green synthesis to produce solvent or by ensuring minimal usage of the solvents if it is not green. Green solvent[47,48] becomes an important issue since the 1990s[49], and the ionic liquid was initially projected as the environment-friendly green solvent[50,51]. In the year 2006, the European Union introduced a regulation named Registration, Evaluation, Authorisation, and Restriction of Chemicals (REACH)[52] to tackle the manufacturing and utilization of chemical substances, and their probable impacts on both human health and the environment. Subsequently, new pathways were searched for the synthesis of green solvents[53] to develop green technologies[54]. However, the measurement of the greenness of the solvent depends on many factors as described by Byrne et al.[55]. They proposed that the solvents must pass through two categories of assessment. One is the assessment of the impact of the solvent on environmental, health, and safety (EHS) issues, which is quite obvious and the other is cumulative energy demand (CED). The CED of solvent can be computed by assessing the energy needed to manufacture the solvent, and whether any opportunity is there at end-of-life to recover some of that energy. Solvatochromic data of a large number of green solvents was compiled by Jessop et al.[56] which can act as an important resource for selecting environment-friendly green solvents. Vian et al.[57] and Clarke et al.[58] reported extensive reviews on the greenness of the solvents and also addressed the sustainability issue. They consider several important solvents such as ionic liquids, deep eutectic solvents, supercritical fluids, switchable solvents, bio-based solvents, liquid polymers, and renewable solvents and evaluated their effective greenness. Other than central regulation for evaluating greenness (e.g., REACH), several researchers also explored different methodologies to evaluate the same. Slater et al.[59] developed a method for the estimation of the greenness of environment-friendly several solvents that are being used in a pharmaceutical process. They incorporated various environmental parameters for assessing the greenness of the particular solvent and determined an overall greenness index. Recently, Miquel et al.[60]also reviewed a quantum chemistry-related method based on the Conductor-like Screening Model (COSMO) as a tool for green solvent screening employing thermodynamic performance indicators.

On the other hand, if the required solvent is not green then one should ensure that the quantity of the solvent that will be used for the analytical purpose must be minimal to reduce its effect on the environment. This can be achieved through the process called microextraction, where the volume of the extracting phase is much lesser in comparison to the volume of the sample[61]. Moreover, the microextraction process can also be achieved using the eco-friendly green approach[62]. Microextraction can be broadly categorised as solid-phase microextraction and liquid-phase microextraction. However, single drop microextraction which falls under the liquid phase category is more popular than others because of its inexpensive nature, easy operating protocol, simplicity, and exceptionally low consumption of solvents. An environment-friendly approach for the liquid-phase microextraction of polar and non-polar acids from urine using a LOC device was recently reported by Santigosa et al.[63].

## 2.4 Green Waste Management

Green waste management is a strategic procedure to ensure maximum safety for the environment by reducing the number of toxic residues. To achieve the said goal, minimal use of the solvents has to be ensured to generate minimal waste. For reduction in generated waste, first of all, the preparation stages of the analytical procedure have to be minimized as a large number of stages will require a higher amount of solvents, thus more will be the volume of the generated waste which may be hazardous for the environment. Furthermore, the volume of samples and solvents must also be reduced to diminish the generation of waste. Microextraction techniques can be employed to fulfil the target where a low concentration of organic solvents is consumed.

Secondly, the generated analytical waste needs to be properly treated so that the generated toxicity of the waste can be diminished. The steps of the waste treatment constitute recycling, degradation, and passivation of waste preferably via green strategies[64]. Recycling is the most preferred method to maintain the environment-friendliness of the process. The recycling of analytical waste can be performed either by distillation or by permeation. To separate volatile organic solvents from non-volatile impurities distillation is used, whereas for selective separation of the organic solvent permeation process is used. If recycling is not feasible, then the next option is degradation, in which several techniques are available, namely thermal degradation, oxidative degradation, heterogeneous photocatalytic oxidations, biodegradation, etc. However, before employing any degradation technique the greenness of the method should be evaluated. For example, thermal degradation depends on the combustion process which may originate toxic gases while in biodegradation some biological processes use the natural metabolism of living cells to degrade or transform analytical waste. When either recycling and degradation method are not practicable due to high cost or large, elongated period, then passivation of the generated analytical waste is opted to make the entire process greener. Co-precipitation of metal ions, surfactant-based decontamination, and adsorption on different materials are some commonly used passivation techniques to minimize the potential threats to the environment arising from analytical waste.

## 2.5 Green Power Management

Like any electronic device, LOC also requires power to operate. To manage the power requirement in an environment-friendly manner LOC system should either generate its power for a stand-alone, self-sustainable working environment or LOC should operate with ultra-low power. Moreover, LOC devices are widely used as POC units where self-powering is a vital issue. Now, self-powering can be achieved by some means where natural energy is converted to electrical energy e.g., solar cells. Liu et al.[65] developed miniaturized biological solar cells which can generate electricity on their own, from microbial (cyanobacteria) photosynthetic and respiratory activities all day long, thus providing a green environment-friendly power source with self-sustaining potential (Fig. 5). Contact-induced electrification (triboelectric effect) is another well-accepted method to generate self-powering for LOC devices[66], especially those which are used for POC[67]. A paper-based inexpensive and self-powered electrochemical device was fabricated by Pal et al.[68] where an integrated paper-based triboelectric nanogenerator (TENG) was used to generate power for the analytical process during POC testing. They used two pieces of hydrophobic paper coated with Ni as the two electrodes of the TENG. A self-powered microfluidic transport system was developed by Nei et al.[69]employing the electrowetting technique and TENG by which a maximum 500 mg load can be transported with a velocity of 1 m/s. Khorami et al.[70] designed LOC-based plasmonic tweezers which require ultra-low power for operation. They fabricated quantum cascaded heterostructures (QCHs) made of Al, In, As, and Ga, and the optical transition within QCHs under biasing of the order of mV, acts as a driver for overlying 1D array of graphene-based environment-friendly plasmonic units.

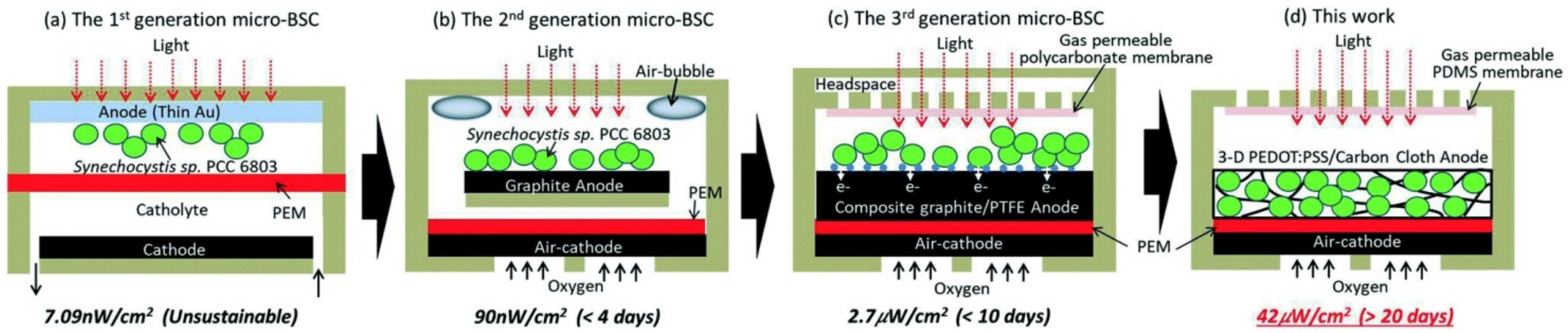

*Fig. 5 (a) Our 1st generation micro-BSC using a face-to-face electrode configuration with a thin gold anode and a liquid electron acceptor. (b) Our 2nd generation micro-BSC using a face-up electrode configuration with an air-bubble trap and a single-chambered air-cathode. (c) Our 3rd generation micro-BSC using a sandwich electrode configuration with a nano-composite graphite/PTFE anode and a microfluidic headspace through a gas permeable polycarbonate membrane. (d) An innovative micro-BSC with a 3-D PEDOT : PSS coated carbon cloth anode and a gas permeable PDMS membrane..* (Reproduced with permission from Lab on a Chip 17, no. 22 (7 November 2017): 3817–25, https://doi.org/10.1039/C7LC00941K)

## 3. Challenges of Commercialization

Microfluidics is capable of creating revolutionary, but rational devices. In microfluidics, a tiny volume of fluids is used on a small microchip for quick delivery of results thus minimizing the complex laboratory procedures. At present, GLOC is the frontline runner among various available LOCs in the market. Moreover, from the aforementioned sections, it seems that the feasibility of fabricating GGLOC is also quite high. Under this optimistic scenario, the scaling of GGLOC from lab to market for commercialization needs to be judged because of the possibility of commercialization if not substantive then pursuing the technology will not be of worth. To commercialize GGLOC, the first barrier obviously will be the firm establishment of its greenness. As of now, no such device is introduced in the market, thus GGLOC has to prove its peerlessness in the context of environment-friendliness. The greenness of the overall analytical process performed by GGLOC can be evaluated in terms of various assessment tools such as the National Environmental Methods Index (NEMI)[71], Green Analytical Procedure Index (GAPI)[72], Analytical Eco-Scale Assessment (ESA)[73], Analytical GREEnness (AGREE)[74], etc. However, assessing the greenness and qualifying in it is not the only issue. There are issues like system integration, economies of scale, and standardization. System integration ever remains an inherent challenge in microfluidics. In general, researchers initially focus on the preparation of different components of GLOC without paying much attention to the integration of these components, which creates the 'chip in a lab' bottleneck. The integration issue must be kept in mind while preparing the components of GLOC and components must be fabricated in such a way that each component is mutually compatible and environment-friendly. Subsequently, in the end, all the components can be easily stitched together to form a functional, fully integrated GGLOC device. Technically complex analyses must run simultaneously on a self-contained portable GGLOC device and should move seamlessly from one step to the next without any hindrance to the successful performance of the chip. The provision for auto-correction to reduce intermediate errors and facility for clear visualization of the result should also be integrated as a 'chip-to-world' interface is crucial for the faithful application of the GGLOC. Lastly, the GGLOC device should be aimed to be integrated with the Internet of Things (IoT)[75] to increase the accessibility of GGLOC devices into the mainstream and to enhance the commercial potential along with market acceptance. An economy of scale is another significant aspect of commercialization. From the perspective of economics, degrees of complexity in the fabrication must be reduced to the bare minimum and from the design phase itself, the researchers should opt for fabrication techniques that are readily up-scalable. The choice of materials is also crucial, predominantly concerning the expenditure and the advantageous intrinsic material properties for the application in question. Toxicity is an inherent hitch when it comes to the application of nanomaterials. A branch of biological science named "nanotoxicology" aims to measure the degree of toxicity caused by nanomaterials. There are various mechanisms like cell autophagy, pyroptosis, apoptosis, and necrosis via which the toxic behaviour can take place. On performing nanotoxicity assessment, it was found that graphene can cause pulmonary toxicity[76], dermal toxicity[77], cardiovascular toxicity[78], reproductive and developmental toxicity[79], hepatotoxicity[80], and ocular toxicity[81].Thus, this is an area of major concern and one need to take a prominent step to reduce the toxicological effect before employing graphene in POC devices. Lastly, standardization is the ultimate step towards the commercialization of any device. For GGLOC the one major issue will be a controllable, reproducible, scalable, and facile synthesis of graphene with specific structure and properties else the standardization of GGLOC will not be feasible. In particular, graphene standardisation is based upon the number of stacking layers and named by their specific terminologies as provided by ISO[82]. However, commercially available graphene is generally ill-defined because of the lack of standardization of these samples, leading to slow-paced commercialization because of confusion at the consumer end. Graphene produced by different methods possesses varied characteristics because of differences in layer number, foreign residues, defect densities, dimensions, metal contamination, etc. which is not reflected in the ISO standards. Thus, clear communication between the graphene suppliers and graphene consumers needs to be developed taking the ISO standardisation into account for proper exploitation of graphene in LOC devices. On the other hand, great diversity in material, for the basic structure of GGLOC allows for greater flexibility conversely, it limits the choice of standard material (e.g., Si in the microelectronic industry) for commercial uptake of the technology. The researchers must strive to set a benchmark where the highest degree of end-result repeatability from chip-to-chip in respect of fabrication tolerances and performance will be ensured to provide the end-user with the most standardized instrumentation possible.

## 4. Conclusion

Greenness is the key to save the mother earth. Thus, every advanced tool and technology must strictly adhere to the rule of greenness. Among the advanced tools, LOC devices set a milestone in the analytical detection of chemical substances and biomolecules. Later on, the inclusion of graphene added an extra feather to the premiumness of LOC. To make the GLOC more sustainable and eco-friendlier, all the components of the GLOC necessitate being fabricated in a green manner. The prospective pathways for manufacturing the components of GGLOC such as base material, core functional material (graphene), and solvents are reviewed here, along with the possibility of employing a green power supply and implementation of a green disposal mechanism. As the potentiality of achieving GGLOC is quite appreciable thus the opportunity of commercialization needs to be evaluated as it will give a thrust in executing more research in this direction. In a sustained production cycle, the first step is fabrication and the next is commercialization. The fabrication step consists of several methods like micro-machining, etching, lithography, thermoforming, moulding, hot embossing, polymer ablation, polymer casting, bonding, etc. These processes are labour-intensive, requiring both expensive equipment and specialized personnel thus act as potential impediments in analytical chemistry LOC applications. However, the latest technological advancement in low-cost 3D printing and laser engraving can be employed to overcome most of these factors. Subsequently, the major issues of commercialization likewise hurdle of integration, maintenance of economies of scale, and the requirement of standardization are discussed. In summary, although there are many impediments to accomplish economical, sustainable, and reproducible, environment-friendly GGLOC, however, it can be anticipated that GLOC will overcome all the blockades in near future employing its enormous potential and will add a green feather to its crown.

---

[1]I.MoserG.JobstE.AschauerP.SvasekM.VarahramG.UrbanV.A.ZaninG.Y.TjoutrinaA.V.ZharikovaT.T.Berezov, 'Miniaturized Thin Film Glutamate and Glutamine Biosensors', *Biosensors and Bioelectronics* 10, no. 6–7 (1995): 527–32, https://doi.org/10.1016/0956-5663(95)96928-R.

[2]G. Czilwik et al., 'Rapid and Fully Automated Bacterial Pathogen Detection on a Centrifugal-Microfluidic LabDisk Using Highly Sensitive Nested PCR with Integrated Sample Preparation', *Lab on a Chip* 15, no. 18 (25 August 2015): 3749–59, https://doi.org/10.1039/C5LC00591D.

[3]Yi Sun et al., 'A Lab-on-a-Chip System with Integrated Sample Preparation and Loop-Mediated Isothermal Amplification for Rapid and Quantitative Detection of Salmonella Spp. in Food Samples', *Lab on a Chip* 15, no. 8 (31 March 2015): 1898–1904, https://doi.org/10.1039/C4LC01459F.

[4]Martin Pumera, 'Nanomaterials Meet Microfluidics', *Chemical Communications* 47, no. 20 (2011): 5671–80, https://doi.org/10.1039/c1cc11060h.

[5]Chun Kiang Chua, Adriano Ambrosi, and Martin Pumera, 'Graphene Based Nanomaterials as Electrochemical Detectors in Lab-on-a-Chip Devices', *Electrochemistry Communications* 13, no. 5 (1 May 2011): 517–19, https://doi.org/10.1016/j.elecom.2011.03.001.

[6]Joydip Sengupta and ChaudheryMustansar Hussain, 'Graphene and Its Derivatives for Analytical Lab on Chip Platforms', *TrAC Trends in Analytical Chemistry* 114 (1 May 2019): 326–37, https://doi.org/10.1016/j.trac.2019.03.015.

[7]GiwanSeo et al., 'Rapid Detection of COVID-19 Causative Virus (SARS-CoV-2) in Human Nasopharyngeal Swab Specimens Using Field-Effect Transistor-Based Biosensor', *ACS Nano* 14, no. 4 (28 April 2020): 5135–42, https://doi.org/10.1021/acsnano.0c02823.

[8]Md Azahar Ali et al., 'Sensing of COVID-19 Antibodies in Seconds via Aerosol Jet Nanoprinted Reduced-Graphene-Oxide-Coated 3D Electrodes', *Advanced Materials* n/a, no. n/a (n.d.): 2006647, https://doi.org/10.1002/adma.202006647.

[9]Rajesh Ahirwar and Amit K. Tripathi, 'E-Waste Management: A Review of Recycling Process, Environmental and Occupational Health Hazards, and Potential Solutions', *Environmental Nanotechnology, Monitoring & Management* 15 (1 May 2021): 100409, https://doi.org/10.1016/j.enmm.2020.100409.

[10]Joydip Sengupta, Arpita Adhikari, and ChaudheryMustansar Hussain, 'Graphene-Based Analytical Lab-on-Chip Devices for Detection of Viruses: A Review', *Carbon Trends* 4 (1 July 2021): 100072, https://doi.org/10.1016/j.cartre.2021.100072.

[35]Zhuang-Jun Fan et al., ‘Facile Synthesis of Graphene Nanosheets via Fe Reduction of Exfoliated Graphite Oxide’, *ACS Nano* 5, no. 1 (25 January 2011): 191–98, https://doi.org/10.1021/nn102339t.

[36]Min Yi et al., ‘A Mixed-Solvent Strategy for Facile and Green Preparation of Graphene by Liquid-Phase Exfoliation of Graphite’, *Journal of Nanoparticle Research* 14, no. 8 (10 July 2012): 1003, https://doi.org/10.1007/s11051-012-1003-5.

[37]Jilei Liu et al., ‘Improved Synthesis of Graphene Flakes from the Multiple Electrochemical Exfoliation of Graphite Rod’, *Nano Energy* 2, no. 3 (1 May 2013): 377–86, https://doi.org/10.1016/j.nanoen.2012.11.003.

[38]Chiranjeevi Srinivasa Rao Vusa, Sheela Berchmans, and Subbiah Alwarappan, ‘Facile and Green Synthesis of Graphene’, *RSC Advances* 4, no. 43 (23 May 2014): 22470–75, https://doi.org/10.1039/C4RA01718H.

[39]Changyan Xu et al., ‘Fabrication and Characteristics of Reduced Graphene Oxide Produced with Different Green Reductants’, *PLOS ONE* 10, no. 12 (14 December 2015): e0144842, https://doi.org/10.1371/journal.pone.0144842.

[40]SrinivasaraoYaragalla et al., ‘Preparation and Characterization of Green Graphene Using Grape Seed Extract for Bioapplications’, *Materials Science and Engineering: C* 65 (1 August 2016): 345–53, https://doi.org/10.1016/j.msec.2016.04.050.

[41]Jiabin Wang, ElifCaliskanSalihi, and LidijaŠiller, ‘Green Reduction of Graphene Oxide Using Alanine’, *Materials Science and Engineering: C* 72 (1 March 2017): 1–6, https://doi.org/10.1016/j.msec.2016.11.017.

[42]Ji-Heng Ding, Hong-Ran Zhao, and Hai-Bin Yu, ‘A Water-Based Green Approach to Large-Scale Production of Aqueous Compatible Graphene Nanoplatelets’, *Scientific Reports* 8, no. 1 (3 April 2018): 5567, https://doi.org/10.1038/s41598-018-23859-5.

[43]Abu Hannifa Abdullah et al., ‘Green Sonochemical Synthesis of Few-Layer Graphene in Instant Coffee’, *Materials Chemistry and Physics* 222 (15 January 2019): 11–19, https://doi.org/10.1016/j.matchemphys.2018.09.085.

[44]BetülGürünlü, ÇiğdemTaşdelenYücedağ, and Mahmut Rahim Bayramoğlu, ‘Green Synthesis of Graphene from Graphite in Molten Salt Medium’, *Journal of Nanomaterials* 2020 (31 January 2020): e7029601, https://doi.org/10.1155/2020/7029601.

[45]Juil Hwang et al., ‘Green Synthesis of Reduced-Graphene Oxide Quantum Dots and Application for Colorimetric Biosensor’, *Sensors and Actuators A: Physical* 318 (1 February 2021): 112495, https://doi.org/10.1016/j.sna.2020.112495.

[46]Qianyu Xu et al., ‘Green Reduction of Graphene Oxide Using Bacillus Sphaericus’, *Journal of Colloid and Interface Science* 605 (1 January 2022): 881–87, https://doi.org/10.1016/j.jcis.2021.07.102.

[47]Christian Capello, Ulrich Fischer, and Konrad Hungerbühler, ‘What Is a Green Solvent? A Comprehensive Framework for the Environmental Assessment of Solvents’, *Green Chemistry* 9, no. 9 (28 August 2007): 927–34, https://doi.org/10.1039/B617536H.

[48]Katharina Häckl and Werner Kunz, ‘Some Aspects of Green Solvents’, *ComptesRendusChimie*, Workshop on alternative solvents for synthesis, extraction, purification, and formulation (WAS 2017) / Workshop international sur les solvantsalternatifs pour la synthèse, l’extraction, la purification et la formulation (WAS 2017) Lyon (France), 27–29 September 2017 / Lyon (France), 27–29 septembre 2017, 21, no. 6 (1 June 2018): 572–80, https://doi.org/10.1016/j.crci.2018.03.010.

[49]J. Sherman et al., ‘Solvent Replacement for Green Processing.’, *Environmental Health Perspectives* 106, no. suppl 1 (1 February 1998): 253–71, https://doi.org/10.1289/ehp.98106s1253.

[50]Martyn J. Earle and Kenneth R. Seddon, ‘Ionic liquids. Green solvents for the future’, *Pure and Applied Chemistry* 72, no. 7 (1 January 2000): 1391–98, https://doi.org/10.1351/pac200072071391.

[51]Hua Zhao, Shuqian Xia, and Peisheng Ma, ‘Use of Ionic Liquids as “Green” Solvents for Extractions’, *Journal of Chemical Technology & Biotechnology* 80, no. 10 (2005): 1089–96, https://doi.org/10.1002/jctb.1333.

[52]‘Regulation (EC) No 1907/2006 of the European Parliament and of the Council of 18 December 2006 Concerning the Registration, Evaluation, Authorisation and Restriction of Chemicals (REACH), Establishing a European Chemicals Agency, Amending Directive 1999/45/EC and Repealing Council Regulation (EEC) No 793/93 and Commission Regulation (EC) No 1488/94 as Well as Council Directive 76/769/EEC and Commission Directives 91/155/EEC, 93/67/EEC, 93/105/EC and 2000/21/EC’, Pub. L. No. 32006R1907, 396 OJ L (2006), http://data.europa.eu/eli/reg/2006/1907/oj/eng.

[53]Philip G. Jessop, ‘Searching for Green Solvents’, *Green Chemistry* 13, no. 6 (7 June 2011): 1391–98, https://doi.org/10.1039/C0GC00797H.

[54]Marina CvjetkoBubalo et al., ‘Green Solvents for Green Technologies’, *Journal of Chemical Technology & Biotechnology* 90, no. 9 (2015): 1631–39, https://doi.org/10.1002/jctb.4668.

[55]Fergal P. Byrne et al., ‘Tools and Techniques for Solvent Selection: Green Solvent Selection Guides’, *Sustainable Chemical Processes* 4, no. 1 (23 May 2016): 7, https://doi.org/10.1186/s40508-016-0051-z.

[56]Philip G. Jessop et al., ‘Solvatochromic Parameters for Solvents of Interest in Green Chemistry’, *Green Chemistry* 14, no. 5 (1 May 2012): 1245–59, https://doi.org/10.1039/C2GC16670D.

[57]MarylineVian et al., ‘Green Solvents for Sample Preparation in Analytical Chemistry’, *Current Opinion in Green and Sustainable Chemistry*, 5 Green Solvents 2017, 5 (1 June 2017): 44–48, https://doi.org/10.1016/j.cogsc.2017.03.010.

[58]Coby J. Clarke et al., ‘Green and Sustainable Solvents in Chemical Processes’, *Chemical Reviews* 118, no. 2 (24 January 2018): 747–800, https://doi.org/10.1021/acs.chemrev.7b00571.

[59]C. Stewart Slater and Mariano Savelski, ‘A Method to Characterize the Greenness of Solvents Used in Pharmaceutical Manufacture’, *Journal of Environmental Science and Health, Part A* 42, no. 11 (31 August 2007): 1595–1605, https://doi.org/10.1080/10934520701517747.

[60]María González-Miquel and Ismael Díaz, ‘Green Solvent Screening Using Modeling and Simulation’, *Current Opinion in Green and Sustainable Chemistry* 29 (1 June 2021): 100469, https://doi.org/10.1016/j.cogsc.2021.100469.

[61]TiinaSikanen et al., ‘Implementation of Droplet-Membrane-Droplet Liquid-Phase Microextraction under Stagnant Conditions for Lab-on-a-Chip Applications’, *Analytica Chimica Acta* 658, no. 2 (25 January 2010): 133–40, https://doi.org/10.1016/j.aca.2009.11.002.

[62]Carene Picot-Allain et al., ‘Conventional versus Green Extraction Techniques — a Comparative Perspective’, *Current Opinion in Food Science* 40 (1 August 2021): 144–56, https://doi.org/10.1016/j.cofs.2021.02.009.

[63]Elia Santigosa et al., ‘Green Microfluidic Liquid-Phase Microextraction of Polar and Non-Polar Acids from Urine’, *Analytical and Bioanalytical Chemistry* 413, no. 14 (1 June 2021): 3717–23, https://doi.org/10.1007/s00216-021-03320-9.

[64]Salvador Garrigues, Sergio Armenta, and Miguel de la Guardia, ‘Green Strategies for Decontamination of Analytical Wastes’, *TrAC Trends in Analytical Chemistry*, Green analytical chemistry, 29, no. 7 (1 July 2010): 592–601, https://doi.org/10.1016/j.trac.2010.03.009.

[65]Lin Liu and Seokheun Choi, ‘Self-Sustainable, High-Power-Density Bio-Solar Cells for Lab-on-a-Chip Applications’, *Lab on a Chip* 17, no. 22 (7 November 2017): 3817–25, https://doi.org/10.1039/C7LC00941K.

[66]Li Zheng et al., ‘Self-Powered Electrostatic Actuation Systems for Manipulating the Movement of Both Microfluid and Solid Objects by Using Triboelectric Nanogenerator’, *Advanced Functional Materials* 27, no. 16 (2017): 1606408, https://doi.org/10.1002/adfm.201606408.

[67]Arpita Adhikari and Joydip Sengupta, ‘Chapter 8 - Nanogenerators: A New Paradigm in Blue Energy Harvesting’, in *Nano Tools and Devices for Enhanced Renewable Energy*, ed. Sheila Devasahayam and ChaudheryMustansar Hussain, Micro and Nano Technologies (Elsevier, 2021), 171–93, https://doi.org/10.1016/B978-0-12-821709-2.00004-9.

[68]Aniket Pal et al., ‘Self-Powered, Paper-Based Electrochemical Devices for Sensitive Point-of-Care Testing’, *Advanced Materials Technologies* 2, no. 10 (2017): 1700130, https://doi.org/10.1002/admt.201700130.

[69]JinhuiNie et al., ‘Self-Powered Microfluidic Transport System Based on Triboelectric Nanogenerator and Electrowetting Technique’, *ACS Nano* 12, no. 2 (27 February 2018): 1491–99, https://doi.org/10.1021/acsnano.7b08014.

[70]Ali Asghar Khorami, Mohammad Kazem MoravvejFarshi, and Sara Darbari, ‘Ultralow-Power Electrically Activated Lab-on-a-Chip Plasmonic Tweezers’, *Physical Review Applied* 13, no. 2 (26 February 2020): 024072, https://doi.org/10.1103/PhysRevApplied.13.024072.

[71]‘National Environmental Methods Index’, accessed 24 October 2021, https://www.nemi.gov/home/.

[72]J. Płotka-Wasylka, ‘A New Tool for the Evaluation of the Analytical Procedure: Green Analytical Procedure Index’, *Talanta* 181 (1 May 2018): 204–9, https://doi.org/10.1016/j.talanta.2018.01.013.

[73]Agnieszka Gałuszka et al., ‘Analytical Eco-Scale for Assessing the Greenness of Analytical Procedures’, *TrAC Trends in Analytical Chemistry* 37 (1 July 2012): 61–72, https://doi.org/10.1016/j.trac.2012.03.013.

[74]Francisco Pena-Pereira, Wojciech Wojnowski, and Marek Tobiszewski, ‘AGREE—Analytical GREEnness Metric Approach and Software’, *Analytical Chemistry* 92, no. 14 (21 July 2020): 10076–82, https://doi.org/10.1021/acs.analchem.0c01887.

[75]Yuan Jia et al., ‘Paper-Based Graphene Oxide Biosensor Coupled with Smartphone for the Quantification of Glucose in Oral Fluid’, *Biomedical Microdevices* 20, no. 4 (12 October 2018): 89, https://doi.org/10.1007/s10544-018-0332-2.

[76]Qinglin Hu et al., ‘Effects of Graphene Oxide Nanosheets on the Ultrastructure and Biophysical Properties of the Pulmonary Surfactant Film’, *Nanoscale* 7, no. 43 (29 October 2015): 18025–29, https://doi.org/10.1039/C5NR05401J.

[77]Marco Pelin et al., ‘Differential Cytotoxic Effects of Graphene and Graphene Oxide on Skin Keratinocytes’, *Scientific Reports* 7, no. 1 (12 January 2017): 1–12, https://doi.org/10.1038/srep40572.

[78]ManjunathaBangeppagari et al., ‘Graphene Oxide Induces Cardiovascular Defects in Developing Zebrafish (Danio Rerio) Embryo Model: In-Vivo Toxicity Assessment’, *Science of The Total Environment* 673 (10 July 2019): 810–20, https://doi.org/10.1016/j.scitotenv.2019.04.082.

[79]Shun Xu, Zheyu Zhang, and Maoquan Chu, ‘Long-Term Toxicity of Reduced Graphene Oxide Nanosheets: Effects on Female Mouse Reproductive Ability and Offspring Development’, *Biomaterials* 54 (1 June 2015): 188–200, https://doi.org/10.1016/j.biomaterials.2015.03.015.

[80]Walaa Nasr Abd El Mohsen Mohamed et al., ‘Evaluation of Genotoxic and Hepatotoxic Effects of Graphene Oxide Nanosheets in Male Albino Mice -’, *THE EGYPTIAN JOURNAL OF EXPERIMENTAL BIOLOGY (Zoology)* 13, no. 1 (2017): 43–53.

[81]Wei Wu et al., ‘Evaluation of the Toxicity of Graphene Oxide Exposure to the Eye’, *Nanotoxicology* 10, no. 9 (20 October 2016): 1329–40, https://doi.org/10.1080/17435390.2016.1210692.

[82]‘ISO/TS 80004-13:2017(En), Nanotechnologies — Vocabulary — Part 13: Graphene and Related Two-Dimensional (2D) Materials’, accessed 9 March 2022, https://www.iso.org/obp/ui/#iso:std:iso:ts:80004:-13:ed-1:v1:en.